\documentclass[12pt]{article}

\usepackage{scicite}
\usepackage{graphicx}%
\usepackage{multirow}%
\usepackage{amsmath,amssymb,amsfonts}%
\usepackage{amsthm}%
\usepackage{mathrsfs}%
\usepackage[title]{appendix}%
\usepackage{xcolor}%
\usepackage{textcomp}%
\usepackage{manyfoot}%
\usepackage{booktabs}%
\usepackage{algorithm}%
\usepackage{algorithmicx}%
\usepackage{algpseudocode}%
\usepackage{listings}%

\usepackage{times}

\newenvironment{sciabstract}{%
\begin{quote} \bf}
{\end{quote}}

\newcounter{lastnote}

\title{All-optical Fourier neural network using partially coherent light}

\author
{Jianwei Qin$^{1,\dag}$, Yanbing Liu$^{2,\dag}$, Yan Liu$^{1,\dag}$, Xun Liu$^{3}$, Wei Li$^{3,\ast}$, Fangwei Ye$^{1,4,\S}$\\
\\
\normalsize{$^{1}$School of Physics and Astronomy, Shanghai Jiao Tong University, }\\
\normalsize{Shanghai 200240, China}\\
\normalsize{$^{2}$School of Electronic Engineering, Beijing University of Posts and Telecommunications,}\\
\normalsize{Beijing,100876,China}\\
\normalsize{$^{3}$Beijing Institute of Space Mechanics and Electricity,China Academy of Space Technology, }\\
\normalsize{Beijing,100094,China}\\
\normalsize{$^{4}$School of Physics, Chengdu University of Technology, }\\
\normalsize{Chengdu, 610059, China }\\
\\
\normalsize{$^\ast$E-mail: wei$\_$li$\_$bj@163.com}\\
\normalsize{$^\S$E-mail: fangweiye@sjtu.edu.cn}\\
\normalsize{$^\dag$Equally contribution}
}

\date{}

\begin{document} 


\baselineskip24pt


\maketitle


\begin{sciabstract}
 Optical neural networks present distinct advantages over traditional electrical counterparts, such as accelerated data processing and reduced energy consumption. While coherent light is conventionally employed in optical neural networks, our study proposes harnessing spatially incoherent light in all-optical Fourier neural networks. Contrary to natural predictions of declining target recognition accuracy with increased incoherence, our experimental results demonstrate a surprising outcome: improved accuracy with incoherent light. We attribute this enhancement to spatially incoherent light's ability to alleviate experimental errors like diffraction rings and laser speckle. Our experiments introduced controllable spatial incoherence by passing monochromatic light through a spatial light modulator featuring a dynamically changing random phase array. These findings underscore partially coherent light's potential to optimize optical neural networks, delivering dependable and efficient solutions for applications demanding consistent accuracy and robustness across diverse conditions, including on-chip optical computing, photonic interconnects, and reconfigurable optical processors.
\end{sciabstract}

\section{Introduction}
All-optical neural networks (AONNs) utilize optical signals for data processing and computation, offering significant advantages over traditional electrical neural networks. The high propagation speed and transmission bandwidth of optical signals enable faster data processing, making AONNs particularly effective for the real-time processing of large datasets and neural network training. Additionally, photon transmission in AONNs is free from the electrical resistance and thermal dissipation found in electrical networks, resulting in lower energy consumption.
AONNs typically perform spatial light computation through coherent diffraction effects. For example, diffraction-deep-neural networks (D2NNs)\cite{lin2018all,yan2019fourier,liu2022programmable,li2021spectrally,zhou2021large,li2024nonlinear,icsil2024all,luo2019design, rahman2021ensemble, luo2022computational} use the spatial adjustment of coherent light field phases to generate complex diffraction effects for optical computation. In image classification tasks, images input via coherent light sources pass through multiple phase masks, and the output light field intensity distribution is used for classification. D2NNs generally require several diffraction phase plates for image recognition. Another efficient architecture is the all-optical Fourier neural networks (AFNNs)\cite{bernstein2023single,liu2024optical,miscuglio2020massively,liu2024towards,yan2019fourier,zuo2019all}, which uses a spatial light modulator to adjust the phase spectrum of the input image light field at the Fourier plane of an optical 4F system. Classification is based on the light intensity distribution of the output field. The simpler and more adjustable optical path structure of the AFNNs, coupled with the coherence of the light field, ensures its high precision.

However, the reliance of AONNs on the output light intensity makes them highly sensitive to experimental errors in the distribution of the light field. Previous experimental studies on AONNs have employed either fully coherent \cite{lin2018all,zuo2019all,spall2022hybrid,zhou2020situ,chen2023deep,zhang2021optical,xu2021optical,li2021spectrally,montes2024fundamentals,zhou2021large,zuo2019all,zhang2024memory,cheng2024multimodal,shen2017deep,li2024nonlinear,kulce2021all,icsil2024all,rahman2021ensemble, fang2024orbital} or fully incoherent \cite{wang2023image,rahman2023universal,fei2023zero} illumination methods. AONNs that are based on coherent light are highly susceptible to phase noise, such as diffraction noise from impurities in the optical system \cite{martin2012noise,godard2012noise} and laser speckle \cite{skipetrov2010noise,volker2005laser,dainty2013laser}. Incoherent light sources, on the other hand, can mitigate the impacts of phase noise, but their accuracy were predicted in a few theoretical studies to fall below that of coherent light \cite{filipovich2024role,filipovich2023diffractive,kleiner2024coherence}. Hence, the question arises: can an all-optical neural network be designed to reduce experimental errors stemming from the coherence of light while maintaining high accuracy?

It is commonly accepted that partially incoherent light favors the suppression of phase noise such as observed in optical holography and interferometry \cite{peng2021speckle,brown1958interferometry,clark2012high,bourassin2015partially,shi2023super,tang2023active}. In this work, we introduce the partially incoherent light into the optical neural networks, and, specifically, develop an AFNN based on partially coherent illumination. Consistent with  previous theoretical studies\cite{filipovich2024role,filipovich2023diffractive,kleiner2024coherence}, our numerical simulations indeed show a decrease in target recognition accuracy with increasing incoherence of the illumination light. Nevertheless, contrary to this numerical result, our experimental observations indicate enhanced accuracy in target recognition with the partially-coherent AFNN compared to its coherent counterparts. We attribute this enhanced performance to the ability of the incoherent light to be insensitive to the phase noise from environmental instability and experimental impurities that are unavoidably present in actual scenarios. Our experimental setup allows for adjustable degrees of spatial incoherence, enabling us to study the characteristics and performance of AFNN under varying degrees of incoherence. The incoherence thus emerges as a novel degree of freedom in optical computation and networks, enhancing the experimental accuracy and robustness of AONNs while driving advancements in on-chip optical computing, photonic interconnects, and reconfigurable architectures. By harnessing spatially partially-coherent light, these systems can suppress coherence-induced noise, improve signal stability, and enable more reliable, scalable implementations of optical neural networks on integrated photonic platforms. This advancement holds promise for the integration and implementation of spatial light computing in realistic real-world conditions, where system tolerance unavoidably plays a crucial role\cite{mengu2020misalignment,mengu2020scale}.

\section{Architecture of partially-coherent all-optical Fourier neural network}

\begin{figure}[h!]
\centering
\includegraphics[scale=1.5]{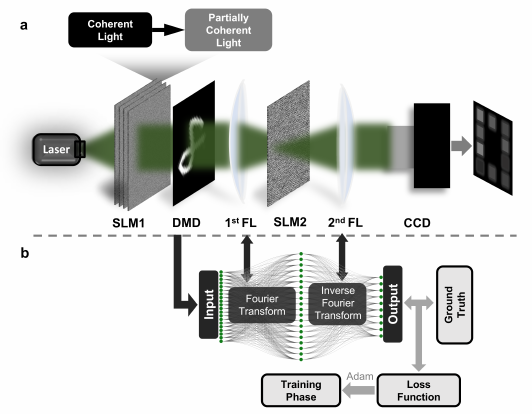}
\caption{(a) Schematic diagram of partially-coherent AFNN architecture. DMD: digital micromirror device loaded with input images; SLM1: spatial light modulator loaded with time-varying random phase map; SLM2: spatial light modulator loaded with a trained phase map; FL: Fourier lens employed as the device for the Fourier transform of the light field. CCD: charge-coupled device employed to capture the output light field.} (b) Flow chart of the training process. The loss function is obtained from the comparison between the output light intensity distribution and the ground truth, and it is used to train the phase map loaded onto SLM2 using the Adam optimization method. Note that the physical transformation of the 4F system is also used for back-propagation training of the AFNN.

\label{fig2}
\end{figure}

Figure \ref{fig2} (a) illustrates the architecture of a partially-coherent AFNN. This system begins with a SLM (SLM1) that rapidly switches a random phase pattern to convert coherent laser light into partially coherent light, serving as the illumination source. A DMD is then used to load the input images for classification. The partially coherent light field passes through a 4F system composed of two Fourier lenses. At the Fourier plane of this system, the second SLM modulates the spatial spectrum's phase of the input light field with phase patterns trained for specific incoherence degrees. The resulting output light field is captured by a CCD, and classification is performed based on the light intensity in the classification region of the output light field.

To be more specific, the light field transmission in the partially coherent AFNN 
follows this process: Coherent light from the laser source, with amplitude $E_0$, passes through the first SLM,  where it acquires a spatially random phase $\phi_{x,y}^{rand}$, 
resulting in $E_1(x,y)=E_0\cdot e^{\phi_{(x,y)}^{rand}}$. By rapidly switching the random phase over time, the light is modulated into a partially coherent state. To simulate the propagating of this partially coherent light field, we treat it as the ensemble average of the propagation results of multiple coherent light fields, each subject to a different initial random phase modulation $\phi_{x,y}^{rand}$. The propagation of each individual light field $E_1$ can be simulated using the Fresnel diffraction integral, 
\begin{equation}
    E_2(x,y)=E_1(x,y)\cdot t(x,y)
\end{equation}
where $t(x,y)$ represents the grayscale of the input image (in our case, handwritten numbers). The light field $E_2$ enters into a 4F system, which consists of two Fourier lenses (each with focal length $f$) and a SLM with a phase distribution $\Psi(k_x,k_y)$) positioned at the Fourier plane of the first lens. The 4F system thus transforms $E_2$ into  $E_3$ as follows, 
\begin{equation}
    E_3(x',y')=\mathscr{F}^{-1}[\mathscr{F}(E_2(x,y)) \cdot e^{i\Psi(k_x,k_y)}]
\end{equation}
where $\mathscr{F}$ denotes the Fourier transform. We simulated the output light fields $E_3(x', y')$ for 100 different random phase modulations of the initial coherent light $E_1(x, y)$, and by performing an ensemble average of these simulations, we obtained the output light intensity $I_{out}$ for the partially coherent light, calculated as $I_{out}(x,y)=\left\langle |E_3(x,y)|^2 \right\rangle$. Further details regarding the propagation of the light field within the setup can be found in the supplementary materials\cite{SM}.

We used the MNIST handwritten digit dataset for training in the image classification task. The training process for the partially coherent AFNN, as illustrated in Fig.\ref{fig2} (b), involves the following steps: (i) \textbf{Output mapping}. The output light field $ E_{out}(x, y) $ is divided into 10 classification spatial regions, each representing to a digit from 0 to 9. The average light intensity within each region $\Omega_i$ is recorded as an array, $(I_0,I_2,...,I_{9})$, where $ I_i=\int_{\Omega_i}I_{out}(x,y) dxdy$. These intensities are then converted into classification probabilities $P_i$ using the softmax function, $P_i=e^{I_i}/\sum_{i=0}^9 e^{I_i}$.
(ii) \textbf{Loss function.} During training, the classification probabilities are compared with the ground truth labels. The network's loss function is defined using cross-entropy: $ L = -\sum_{i=0}^9 y_i \cdot \log(P_i) $ where $y_i$ is a one-hot encoded vector representing the correct class (with $y_i = 1$ for the correct class and $y_i = 0$ otherwise).
(iii ) \textbf{Backpropagation to train phase patterns}. The loss function is propagated back to the phase patterns on the second SLM using gradient descent, with the Adam optimizer computing the gradients and updating the training parameters.
(iv) \textbf{Iterative training}. These steps are repeated iteratively, refining the phase patterns to improve classification accuracy until convergence is achieved. The final result is the optimal phase pattern $\Psi(k_x,k_y)$.

\section{Experimental results}

\textbf{Generation and measurement of partially coherent light} Generating controllable partially-coherent light as an illumination source is essential for partially-coherent AFNN. We used a spatial light modulator (SLM) as a controllable dynamic scattering medium to generate partially-coherent light \cite{cai2017generation}. In the experiment, a 532 nm wavelength laser source illuminates the SLM, which is loaded with time-varying random phase maps. These random phase patterns in SLM have a range of $[(1-\delta)\pi,(1+\delta)\pi]$, where $\delta$ represents the modulation depth ($0<\delta<1$). By rapidly switching the random phase maps on the SLM (at a 20 ms interval), the incident coherent light is converted into partially-coherent light. Adjusting the modulation depth $\delta$ on the SLM allows control over the degree of incoherence $\mu_{light}=1-\mathrm{sinc}^2(\delta)$. More detail can be found in supplementary material\cite{SM}.

\begin{figure}[h!]
\centering
\includegraphics[scale=1.5]{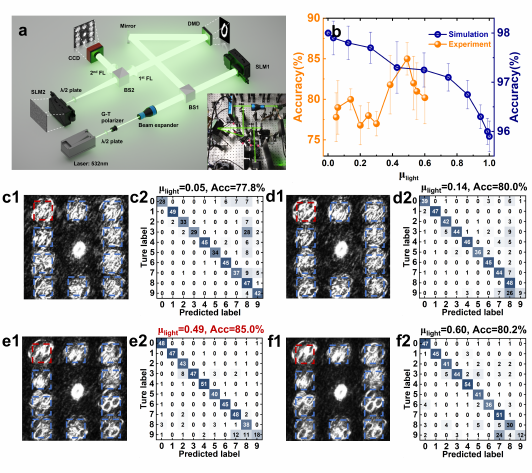}
\caption{(a) Experimental schematic of the partially-coherent AFNN with an inset showing a photo of the experimental setup. (b) Simulated and experimental classification accuracy of the partially-coherent AFNN on the MNIST dataset under varying degrees of incoherence, both averaged 10 independent realizations in the phase map loaded onto SLM2. (c1-f1) Output light fields captured by the CCD at incoherence degrees of 0.05, 0.14, 0.49, and 0.60, respectively. The classification is based on the mean light intensity within boxed regions, and  
the region of maximum light intensity determines the classification result (the red box for the specific test shown). (c2-f2) Confusion matrices depicting the experimental inference results of the partially-coherent AFNN at incoherence degrees of 0.05, 0.14, 0.49, and 0.60, respectively.}
\label{fig3}
\end{figure}

\textbf{Partial coherence enhanced AFNN.} Based on the architecture shown in Fig.\ref{fig2} (a), we constructed the experimental optical setup for the partially-coherent AFNN, as depicted in Fig.\ref{fig3} (a). A 532 nm laser served as the light source and it was converted into linearly polarized light using a half-wave plate and a Glan-Taylor polarizer, optimizing the modulation efficiency of the SLM. We used the Holoeye-LETO-80R SLM, which has a resolution of $1920\times1200$, a pixel size of $8~\mu \text{m}\times 8~\mu \text{m}$, a fill factor of 95\%, and a phase modulation range of 0 to $2\pi$ (corresponding to grayscale values of 0 to 255). The first SLM, responsible for converting coherent light into partially coherent light, had an active area of $15.4 ~\text{mm} \times 9.6~\text{mm}$, exceeding the beam size (approximately $8.3~$mm radius) to ensure full beam modulation. However, the minimum unit size for random phase modulation, $8~\mu \text{m}\times 8~\mu \text{m}$ , was larger than the ideal infinitesimal size, resulting in a lower degree of incoherence in the experiment compared to simulations. The DMD used to load input images was a Texas Instruments DLPC900, with a resolution of $1920\times1080$ and a pixel size of $7.56~\mu \text{m}\times 7.56~\mu \text{m}$. In the 4F system used for image classification, both Fourier lenses had a focal length of $15~$mm, and the second SLM loaded phase maps trained for specific degree of incoherence. The CCD used to capture output images had a maximum light intensity threshold of $13.8~ \text{mW/m}^2$, truncating any intensity beyond this limit. For a given degree of incoherence, an optimal phase map is obtained through the aforementioned training process and loaded on SLM2. In the numerical simulations, we used 10,000 test images to compute the test accuracy of the partially-coherent AFNNs. In the experiments, to minimize the impact of laser output intensity fluctuations on the results, we used 500 test images from the MNIST dataset to verify the classification accuracy of the partially-coherent AFNNs under different degrees of incoherence\cite{SM}.

As shown by the blue curve in Fig.~\ref{fig3}(b), our simulations indicate that increasing incoherence typically leads to a decrease in network accuracy, with 98.0\% accuracy at full coherence ($\mu_{light}=0$) and 95.9\%  accuracy at full incoherence ($\mu_{light}=1.0$). However, in the experiment, coherent light introduces detrimental effects such as diffraction rings and laser speckle, resulting in a lower accuracy of 77.8\% under almost fully coherent conditions ( Fig.\ref{fig3}, c1 and c2). Remarkably, by introducing partial incoherence, we observed an improvement in experimental accuracy. At a moderate incoherence degree of 0.14, the experimental accuracy increased to 80.0\% (Fig.~\ref{fig3}, d1 and d2), and the highest accuracy of 85.0\% was achieved at an incoherence degree of 0.49 (Fig.~\ref{fig3}, e1 and e2). 
However, it should be noted that further increasing the level of incoherence beyond the optimal point led to a decrease in accuracy (Fig.~\ref{fig3}, f1 and f2). The experimental trend of classification accuracy as a function of incoherence is presented by the orange curve in Fig.~\ref{fig3} (b).

\begin{figure}
\centering
\includegraphics[scale=1.5]{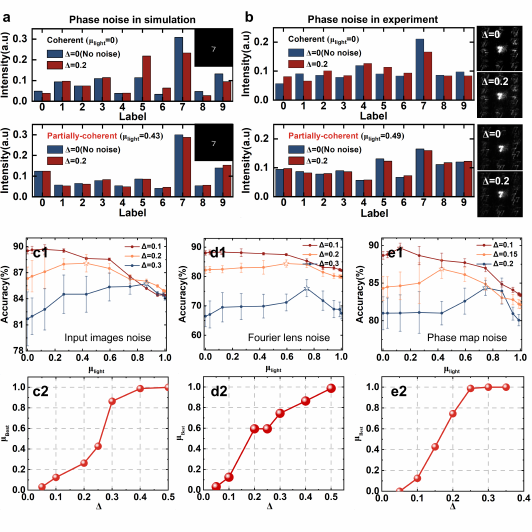}
\caption{(a-b)Under both noisy (red bars) and noise-free (blue bars) conditions, the relative intensities of the output light fields in different classification regions for the coherent AFNN (top) and the relative intensities of the output light fields in different classification regions for the partially-coherent AFNN (bottom). (c1-e1) In the different amplitudes of phase noise $\Delta$, the simulated classification accuracy of AFNN as a function of incoherence degree $\mu_{light}$. (c2-e2)  the degree of incoherence $\mu_{best}$ that yields the highest classification accuracy under different amplitudes of phase noise. The phase noise is respectively introduced in input images (c1,c2), Fourier lens (d1,d2), and phase map (e1,e2).}

\label{fig4}
\end{figure}

The counterintuitive result seen from the experimental curve of Fig.~\ref{fig3} (b)-where accuracy improves as incoherence increases prior to the optimal incoherent point-originate from the effectiveness of partially coherent light in mitigating the adverse effects associated with coherent light. 
Indeed, the adverse diffraction effects are generally induced by inevitable phase noise in experiment setups. By comparing the output light fields of the AFNN in the presence and absence of phase noise, we can validate the suppressive impact of partially coherent light on such noise. Fig.\ref{fig4}(a) and (b) illustrate the variations in light intensity within the classification regions resulting from phase noise introduced into the network (specifically on SLM2) in simulations and actual experiments, respectively. For the fully coherent AFNN, the incorporation of randomly distributed phase noise within the interval $[-\Delta \cdot \pi, \Delta \cdot \pi]$ significantly distorts the light intensity distribution of the output light field (depicted by red bars) when compared to the ideal noise-free environment (blue bars), even when the phase noise is weak $(\Delta = 0.2)$. In contrast, the partially-coherent AFNN exhibits no significant changes in the light intensity distribution of the output light field when phase noise is introduced. This result corroborates that utilization of partially coherent light in an AFNN effectively diminishes the impact of diffraction effects caused by experimental phase noise on the output light field, thereby improving  experimental accuracy. 

An improvement in classification accuracy with increasing incoherence is also observed when phase noise is applied to different components of the network, as shown in the simulation results of Fig.\ref{fig4} (c1-e1). These networks were trained under noise-free conditions with varying degrees of incoherence, and after the training the networks were tested by applying the phase noise to the input image (DMD) [Fig.\ref{fig4} (c1)], the Fourier lens [Fig.\ref{fig4} (d1)], or the phase map (SLM2) [Fig.\ref{fig4} (e1)].  As shown in Fig.\ref{fig4} (c1), when phase noise with \(\Delta = 0.3\) was applied to the input images, the coherent AFNN (\(\mu_{light} = 0\)) achieved a classification of 81.70\%, whereas the partially-coherent AFNN attained the highest accuracy of 85.42\% at \(\mu_{light} = 0.74\). For phase noise with \(\Delta = 0.2\), the partially-coherent AFNN reached its highest accuracy at \(\mu_{light} = 0.42\). Fig.\ref{fig4} (c2) illustrates the degrees of incoherence ($\mu_{best}$) corresponding to the highest classification accuracy for the partially-coherent AFNN under varying phase noise amplitudes. The results indicate that as the phase noise amplitude \(\Delta\) increases, the degree of incoherence \(\mu_{best}\) required to achieve the highest accuracy also rises. When \(\Delta \geq 0.4\), the AFNN illuminated by fully incoherent light (\(\mu_{best} = 1.0\)) exhibits the highest classification accuracy.

Similarly, for phase noise introduced on the Fourier lens [Fig.\ref{fig4} (d1)] and the phase map [Fig.\ref{fig4} (e1)], the classification accuracy of the partially-coherent AFNN also surpasses that of the coherent AFNN. Moreover, as the noise amplitude \(\Delta\) increases, the incoherence degree required to achieve the highest accuracy also grows (detailed discussions are provided in the supplementary materials\cite{SM}). Therefore, in systems with larger phase noise amplitudes, the use of optical neural networks with a higher degree of incoherence becomes more advantageous.

\textbf{Incoherence tolerance in partially coherent AFNNs.}
Now, suppose the phase map, which has been trained under illumination by partially coherent light with a degree of incoherence (denoted as $\mu_{train}$ in following), is loaded onto SLM2. This trained AFNN is expected to perform optimally when the illumination light has the same degree of incoherence as in the training process. However, in practical applications, the
degree of incoherence of the actual illumination light, denoted as $\mu_{light}$, may differ from $\mu_{train}$. This raises the question: how does the performance of the partially coherent AFNN, trained with the light of incoherence $\mu_{train}$, change when in actual applications the incoherence of the illumination light, $\mu_{light}$, differs?

\begin{figure}[h!]
\centering
\includegraphics[scale=1.5]{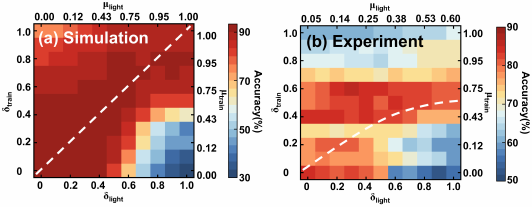}
\caption{(a) The simulated inference accuracy of the partially-coherent AFNN, assessed with a phase map trained for incoherence degree of \(\mu_{train}\) (random phases modulation depth \(\delta_{train}\)), but illuminated by partially coherent light with incoherence degree of \(\mu_{light}\) (random phase modulation depth \(\delta_{light}\)). (b) The same as (a), but for the experimentally measured inference accuracy. The white dashed lines represent points where $\mu_{light}=\mu_{train}$.
}
\label{fig5}
\end{figure}

To answer this question, we study the classification accuracy of systems trained with a certain degree of incoherence $\mu_{train}$, but illuminated with light of incoherence $\mu_{light}$ that may occur in real applications. The results are shown in Fig.\ref{fig5}, with simulations in (a) and experimental data in (b), both plotted on a 2D plane with axes $\mu_{light}$ and $\mu_{train}$ (or equivalently $\delta_{light}$ and $\delta_{train}$, recalling that the modulation depth of the phase determines the degree of incoherence). As expected, the highest classification accuracy is generally achieved when the incoherence of the illumination light matches the training condition ($\mu_{light} = \mu_{train}$), as indicated by the white dashed line in Fig.\ref{fig5}. In the simulation, when the incoherence of the illumination light, $\mu_{light}$, deviates from that of the trained model, $\mu_{train}$—whether $\mu_{light}$ is higher or lower than $\mu_{train}$—the accuracy decreases, but in notably different ways. For example, when $\mu_{light} > \mu_{train}$, meaning the system is illuminated by light with higher incoherence than it was trained for, the classification accuracy drops sharply as the incoherence mismatch increases. Specifically, for an AFNN trained with fully coherent light ($\mu_{train} = 0$), the simulation results show that the accuracy is 97.2\% when $\mu_{light} = 0$, but drops drastically to 32.4\% when $\mu_{light} = 1.0$. In contrast, when $\mu_{light} < \mu_{train}$, meaning the system is illuminated by light with lower incoherence than it was trained for, the accuracy remains relatively high despite the incoherence mismatch. For an AFNN trained with fully incoherent light ($\mu_{train} = 1.0$), the simulation results show that, while the accuracy is 91.6\% when $\mu_{light} = 1.0$, even when fully coherent light is used for illumination ($\mu_{light} = 0$), the accuracy remains as high as 79.2\%.

The experimental findings reveal a similar trend. When $\mu_{light} > \mu_{train}$, for an AFNN trained with $\mu_{train} = 0$, the accuracy drops from 77.8\% when $\mu_{light} = 0.05$ to 55.9\% when $\mu_{light} = 0.60$. Conversely, when $\mu_{light} < \mu_{train}$, the accuracy remains relatively stable. For example, for an AFNN trained with $\mu_{train} = 0.60$, the experimental results show that the accuracy is 80.2\% when $\mu_{light} = 0.60$, and only slightly decreases to 79.5\% when $\mu_{light} = 0.05$.

\begin{figure}[h!]
\centering
\includegraphics[scale=1.5]{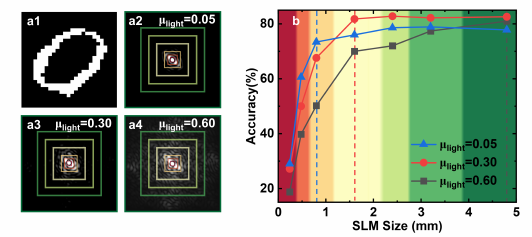}
\caption{(a1) An image of the digit '0' from the MNIST dataset. (a2-a4) Experimentally measured intensity spectra of the image in (a1) when illuminated with partially coherent light, with degree of incoherence set to 0.05, 0.30, and 0.60, respectively. (b) Inference accuracy of the AFNN measured as a function of SLM size, corresponding to the seven square regions indicated in (a2-a4). The dashed vertical lines in (b) indicate the SLM sizes at which the accuracy converges.
}
\label{fig6}
\end{figure}

The ability of an AFNN trained with high incoherence to maintain high accuracy under lower incoherence illumination, while an AFNN trained with low incoherence struggles under highly incoherent illumination, is connected to the phase modulation of the spatial spectrum generated by the input image at the Fourier plane. Typically, the spatial spectrum expands as the incoherence of the illumination light increases. Using Fig.\ref{fig6}  as an illustration for the same input image (a digit “zero”), when lit with low incoherent light ($\mu_{light}=0.05$), the spatial spectrum is mainly concentrated near the center of the Fourier plane (Fig.\ref{fig6}, a2), occupying the third smallest square). In contrast, under a light with $\mu_{light}=0.60$, the spectrum widens significantly, even reaching the largest square (Fig.\ref{fig6}, a4). Therefore, in practical applications, when AFNNs are trained with a certain incoherence light with $\mu_{train}$, using a higher incoherent illumination source ($\mu_{light} > \mu_{train}$) leads to broader coverage of the phase plate at the Fourier plane compared to the trained one, thus resulting in a significant drop in classification accuracy. Conversely, with lower incoherent light ($\mu_{light} < \mu_{train}$), the Fourier spectrum occupies a smaller portion of the already-trained phase plate, thus maintaining relatively high accuracy (though lower than with matched incoherent illumination). This rationale is further supported by Fig.\ref{fig6} (b), where the performance of AFNNs under different incoherent light illuminations is evaluated with varying phase plate sizes (specifically, SLM2 in our setup). The results show that as the size of the phase plate increases, the network accuracy improves for all illumination conditions but saturates at different plate sizes. For instance, with light at $\mu_{light}=0.05$ (blue line in Fig.\ref{fig6} (b)), accuracy saturates at an SLM size of $0.8\times 0.8\text{mm}^2$, as this size is already large enough to modulate all spectral components at the Fourier plane (corresponding to the third smallest box in Fig.\ref{fig6} , a2). In contrast, for $\mu_{light}=0.30$, accuracy convergence requires SLM sizes larger than $1.6\times 1.6\text{mm}^2$ (forth smallest box in Fig.\ref{fig6} , a3). Further,  for $\mu_{light}=0.60$, accuracy stabilizes for SLM sizes exceeding $4.8\times 4.8\text{mm}^2$ (the biggest box in Fig.\ref{fig6} , a4). This highlights that the classification accuracy continues to rise as the SLM size increases to handle higher Fourier components, plateauing once surpassing the spectral range, indicating that further size increases do not impact accuracy.

\section{Conclusion}
In conclusion, we have developed an all-optical Fourier neural network (AFNN) utilizing partially coherent light, significantly improving experimental accuracy compared to fully coherent systems. By adjusting the degree of incoherence in the illumination source, we address coherence-induced experimental errors, which could potentially provide an additional means to optimize optical computation. Our experiments with the MNIST dataset show that partially-coherent AFNN achieves higher accuracy at optimal incoherence degrees, mitigating the detrimental effects of coherent light such as diffraction rings and laser speckles. 

Looking ahead, the capability to fine-tune the degree of spatial incoherence opens up new possibilities to enhance the performance and accuracy of optical neural networks across diverse environments. This advancement not only improves the potential of AONNs but also opens doors to a wider range of applications,  including real-time image classification, biomedical imaging, and optical signal processing, where accuracy and robustness are paramount. Incorporating spatially incoherent illumination into integrated photonic platforms could help mitigate coherence-induced noise in on-chip optical neural networks, improving stability and reliability for chip-scale optical computing and signal processing.

Furthermore, our proposed architecture has the experimental flexibility to straightforwardly include optically nonlinear active functions  (e.g., implemented with optical nonlinear crystals), which is expected to play a vital role in handling more challenging tasks. Finally, our proposed spatially incoherent AONN, combined with temporally incoherent modulation \cite{dong2024partial}, holds promise for enabling integrated photonic chips to operate under spatially and temporally incoherent conditions (i.e., "white light"), including natural light. Optical neural networks effectively operating under natural light scenarios hold substantial promise for practical and robust applications, and our results lay a promising foundation for realizing this vision.

\textbf{note added} After submission \cite{qin2024all}, we became aware of a concurrent and independent study \cite{jia2024partially} that also introduced partial spatial coherence into diffractive optical neural networks, thus breaking the requirement for fully coherent conditions. Both works incorporate partial coherence into optical neural networks; however, we adopt a different architecture by employing an optical Fourier neural network. Furthermore, we reveal a seemingly counterintuitive yet critical phenomenon: partially coherent systems often outperform fully coherent systems in realistic scenarios because they inherently suppress phase noise. Their study\cite{jia2024partially} also observed performance enhancements when using incoherent light, and our work comprehensively investigates this noise-resistant effect, highlighting both the distinct advantages of partial coherence and the broader potential of partially coherent optical neural networks.

\section*{Author contributions}
F.Y. and W.L. initiated and supervised the project. J.Q. designed the research and methodology. J.Q. and Y.L. carried out the simulations and data processing. J.Q. conducted the experiments. Y.L. and.J.Q. developed the experimental system. J.Q, F.Y., Y.L, Y.L, X.L., and W.L. analyzed the results. F.Y. and J.Q. prepared the manuscript with contributions from all authors. All authors participated in discussions regarding the research.

\section*{Data availability}
All the data and methods required to assess the conclusions of this study are provided in the main text. Additional data can be requested from the corresponding author. The codes used in this study rely on standard libraries and scripts, which are publicly accessible in PyTorch.

\section*{Acknowledgements}
F. Y acknowledges the support from Shanghai Outstanding Academic Leaders Plan and Shanghai Leading Talents Program. This work was supported in part by the National Natural Science Foundation of China under grant U23B2011.

\section*{Conflict of interest}
No author has reported any competing interests.

\bibliography{ref}

\bibliographystyle{sciencemag}

\end{document}